\documentclass{article}
\usepackage{spconf,amsmath,graphicx,hyperref,amssymb,color, amsthm}

\title{Robust MMSE Precoding for Out-of-Cluster Interference Mitigation in Cell-Free MIMO Networks}
\name{André R. Flores$^{1}$ and Rodrigo C. de Lamare$^{2,3}$ \vspace{-0.75em} \thanks{This work was partially supported by CAPES, CNPq, FAPESP and FAPERJ 
under grant E-26/205.840/2022 of process SEI-260003/019660/2022.}}
\address{$^{1}$ University of Sao Paulo, Brazil, $^{2}$ PUC-Rio, Rio de Janeiro, Brazil, 
$^{3}$University of York, UK \vspace{-0.5em}}
\begin{document}
\ninept
\maketitle
\begin{abstract}
In this work, we develop a linear robust minimum mean-square error (RMMSE) precoder to mitigate the effects of imperfect channel state information (CSI) and the intra-cluster (ICL) and out-of-cluster (OCL) interference in cell-free (CF) multiple-antenna systems. The proposed precoder includes statistical information of the OCL interference in its derivation, allowing a more effective interference mitigation. An analysis of the sum-rate that can be obtained by the CF system is carried out and an expression quantifying the theoretical gains of mitigating OCL interference are derived. Simulation results corroborate that the proposed RMMSE precoder effectively mitigates ICL and OCL interference.
\end{abstract}
\begin{keywords}
Cell-free, out-of-cluster interference, interference mitigation, robust techniques. \vspace{-0.5em}
\end{keywords}
\section{Introduction}
\label{sec:intro}

Cell-free (CF) and large-scale multiple-input multiple-output (MIMO) systems have attracted a great deal of research interest in the last few years \cite{Elhoushy2021,Ngo2024,mmimo,wence}. In contrast to conventional wireless communications systems, which rely on centralized base stations (BS), CF systems deploy multiple distributed access points (APs) over the geographical area of interest. The distributed infrastructure of CF allows the network to take advantage of the wireless characteristics. As a result, CF-MIMO systems provide a higher throughput per user and better energy efficiency (EE) than conventional systems \cite{Ammar2022,cesg,rra,rrsprec,rlspa}. For this reason, CF MIMO systems have become a potential technology that meets the continuously increasing demands of future wireless communications networks, such as higher data rates and lower latency \cite{Wang2023,cdidd,iddllr}. 

Even with its advantages, CF systems face non-trivial problems such as signaling load, computational cost, imperfect channel state information (CSI) and interference. To deal with the simultaneous transmission  of multiple users employing the same time-frequency resources, CF-MIMO systems implement precoders in the downlink. Specifically, linear precoders, such as the conjugate beamforming (CB) \cite{Nayebi2017}, the zero-forcing (ZF) \cite{Nguyen2017}, and the minimum mean-square error (MMSE) precoders \cite{Palhares2021,iapsprec,rmmsecf}, have attracted the attention of researchers due to their low computational complexity.

Additionally, the deployment of multiple APs results in a computational demanding procedure to estimate the channel gains, increasing also signaling load. For this reason, network-wide (NW) techniques that employ all the APs for simultaneous transmissions to all users are not suitable for practical systems \cite{Bjoernson2020a}. To overcome complexity and signaling problems, CF systems based on clusters of APs and UEs have been proposed \cite{Buzzi2017,Buzzi2020,rscf,srclust,rsa}, avoiding the use of the NW approach. By employing clusters a small set of channel estimates is required to be estimated and conveyed. 

While clustered CF systems deal with scalability and complexity problems, they also introduce out-of-cluster (OCL) interference in addition to intra-cluster (ICL) interference, which degrades the performance of CF systems, limiting their potential. Similar scenarios where interference arrives from outside the network, have been studied. For instance, the suppression of co-channel interference in MIMO was studied in \cite{Saddek2007}, where a leakage-based precoder is implemented. The suppression of interference in multicell scenarios of clustered MIMO networks was considered in \cite{Zhang2009,Zhang2010,Choi2007}. In \cite{Ye2016}, techniques to manage the interference produced by implementing and adding small cells in massive MIMO systems are introduced. Recently, the authors of \cite{Shaik2024} propose algorithms to deal with the out-of-system interference in distributed MIMO networks. In \cite{Ssettumba2025}, an iterative soft ICL and OCL interference cancellation scheme for the uplink of cluster-based CF-MIMO systems was developed. However, the OCL interference in the downlink of CF systems constitutes a different scenario with its own particularities. Consequently, there is an urgent demand for techniques that can mitigate imperfect CSI, ICL and OCL interference in the downlink of CF-MIMO systems. 


In contrast to other works, we propose a robust MMSE (RMMSE) precoding technique to deal with imperfect CSI, ICL and OCL interference in the downlink of CF-MIMO networks. In particular, we develop a RMMSE precoder that considers statistical information about the OCL interference in the robust design. An analysis of the sum-rate of the proposed RMMSE precoder quantifies the gains obtained by accurately modeling ICL and OCL interference. Numerical results illustrate the performance of the proposed RMMSE precoder with ICL and OCL interference mitigation in scenarios of practical interest for CF-MIMO networks. 

The paper is organized as follows. Section 2 introduces the system model. Section 3 details the proposed robust MMSE precoder, whereas Section 4 develops a sum-rate analysis. Section 5 presents and discusses the results, and Section 6 draws the conclusions. \vspace{-0.5em}


\section{System Model}

Let us consider a clustered CF multiple-antenna network, where $M$ APs are deployed and provide service to $K$ users. The APs and the users are separated into $C$ disjoint clusters to lower the signaling load and the computational cost. In particular, the number of APs grouped in cluster $i$ is given by $M_i$. Similarly, $K_i$ is the number of users in cluster $i$. Thus, we have $\sum_{i=1}^C M_i=M$, and $\sum_{i=1}^C K_i=K$. The simultaneous transmission of information in the clusters produces OCL interference. 

Let us denote by $\mathcal{M}^{\left(o\right)}_{i,m}$ the set containing the index of APs in cluster $m$ that produce interference when decoding the information intended for users in cluster $i$. Furthermore,   $M^{\left(o\right)}_{i,m}$ denotes the cardinality of $\mathcal{M}^{\left(o\right)}_{i,m}$. Then,  $M^{\left(o\right)}_{i,m}\leq M_m$. The data intended for the $i$-th cluster are in the transmit vector $\mathbf{x}_{i}\in \mathbb{C}^{M_i}$, which is sent through a flat-fading channel given by $\mathbf{G}_i^H\in\mathbb{C}^{K_i\times M_i}$. Any other transmit vector, say $\mathbf{x}_m$, produces OCL interference since it arrives at the users in cluster $i$ via the channel matrix $\mathbf{G}^{H}_{i,m}\in\mathbb{C}^{K_i\times M_m}$. In other words, $\mathbf{G}_{i,m}^H$ is the channel of OCL interference from all APs in cluster $m$ to the user in cluster $i$. The received signal at the users in the $i$-th cluster is given by
\begin{equation}
\mathbf{y}_i=\mathbf{G}^H_i\mathbf{x}_i+\sum\limits_{m=1,m\neq i}^{C}\mathbf{G}^{H}_{i,m}\mathbf{x}_{m}+\mathbf{n}_i,\label{receive vector at cluster i complete}
\end{equation}
where $\mathbf{n}_i\in \mathbb{C}^{K_i}$ denotes the additive white Gaussian noise (AWGN) at the receivers in cluster $i$. Specifically, $\mathbf{n}_i\sim(\mathbf{0},\sigma_n^2\mathbf{I}_{K_i})$.

We can simplify \eqref{receive vector at cluster i complete} using the binary matrix $\mathbf{U}_{i,m}\in^{M_m \times M_{i,m}^{\left(o\right)}}$, which selects the APs in cluster $m$ that generate strong or non-negligible interference in cluster $i$. In other words, $\mathbf{G}^{\left(o\right)^H}_{i,m}=\mathbf{G}^H_{i,m}\mathbf{U}_{i,m}$ where $\mathbf{G}^{\left(o\right)^H}_{i,m}\in\mathbb{C}^{K_i \times M_{i,m}^{\left(o\right)}}$ contains the channel of the most relevant APs in the $m$-th cluster, and therefore outside the cluster $i$. It follows that the received vector at cluster $i$ is given by  
\begin{equation}
\mathbf{y}_i^{\left(o\right)}=\mathbf{G}^H_i\mathbf{x}_i+\sum\limits_{m=1,m\neq i}^{C}\mathbf{G}^{\left(o\right)^H}_{i,m}\mathbf{x}_{i,m}^{\left(o\right)}+\mathbf{n}_i,
\end{equation}
 where the transmit vector $\mathbf{x}_{i,m}^{\left(o\right)}=\mathbf{U}_{i,m}^{\text{T}}\mathbf{x}_m\in\mathbb{C}^{M_{i,m}^{\left(o\right)}}$ contains information intended for users in the $m$-th cluster, but produces relevant interference to users in cluster $i$.  

To generalize the model for all clusters, let us define the channel block diagonal matrix $\mathbf{G}^H=\textrm{diag}\left(\mathbf{G}_1^H,\mathbf{G}_2^H,\cdots, \mathbf{G}_C^H\right)\in\mathbb{C}^{K\times M}$. On the other hand, $\mathbf{\bar{U}}_{i,m}\in^{M_m \times M_{m}}$ is an expanded version of matrix $\mathbf{U}_{i,m}$, which has zero vectors at the positions of the APs that were not selected. By defining $\mathbf{\bar{G}}^{\left(o\right)^H}_{i,m}=\mathbf{G}^H_{i,m}\mathbf{\bar{U}}_{i,m}$, we have
\begin{equation}
    \mathbf{G}^{\left(o\right)^H}=\begin{bmatrix}
\mathbf{0} &\mathbf{\bar{G}}_{1,2}^{\left(o\right)^H} &\mathbf{\bar{G}}_{1,3}^{\left(o\right)^H} &\cdots &\mathbf{\bar{G}}_{1,C}^{\left(o\right)^H}\\
\mathbf{\bar{G}}_{2,1}^{\left(o\right)^H} &\mathbf{0} &\mathbf{\bar{G}}_{2,3}^{\left(o\right)^H} &\cdots &\mathbf{\bar{G}}_{2,C}^{\left(o\right)^H}\\
\vdots &\vdots &\vdots &\ddots &\vdots\\
\mathbf{\bar{G}}_{C,1}^{\left(o\right)^H} &\mathbf{\bar{G}}_{C,2}^{\left(o\right)^H} &\mathbf{\bar{G}}_{C,3}^{\left(o\right)^H} &\cdots &\mathbf{0}
    \end{bmatrix} \in \mathbb{C}^{K\times M},
\end{equation}
Then, the received signal of the whole network is given by
\begin{equation}
\mathbf{y}=\mathbf{G}^H\mathbf{x}+\mathbf{G}^{\left(o\right)^H}\mathbf{x}+\mathbf{n}.
\end{equation}
where $\mathbf{x}=\left[\mathbf{x}_1^H,\mathbf{x}_2^H,\cdots,\mathbf{x}^H_C\right]^H\in\mathbb{C}^{M}$ is the vector that contains the transmitted symbols and $\mathbf{n}\in\mathbb{C}^K$ is the AWGN, which follows a complex Gaussian distribution, i.e., $\mathbf{n}\sim\mathcal{CN}\left(\mathbf{0},\sigma^2_n\mathbf{I}_K\right)$.

The transmit vector $\mathbf{x}$ is obtained as follows. First, the information is modulated into a vector of symbols $\mathbf{s}=\left[\mathbf{s}^H_1,\mathbf{s}^H_2,\cdots,\mathbf{s}_C^H\right]^H\in \mathbb{C}^K$, where $\mathbf{s}_i \in \mathbb{C}^{K_i}$ conveys the information intended for the users in cluster $i$. The components of $\mathbf{s}$ are independent and identically distributed with unit power. Then,  a precoding matrix $\mathbf{P}\in \mathbb{C}^{M\times K}$ maps the symbols to the transmit antennas. Thus, we have
\begin{equation}
    \mathbf{x}=\mathbf{P}\mathbf{s}.
\end{equation}
The system obeys a total transmit power constraint, i.e., $\mathbb{E}\left[\lVert\mathbf{x}\rVert^2\right]=P_t$ and employs the time-division duplexing (TDD) protocol. Therefore, the channel is obtained by employing the reciprocity property. In particular, the coefficient $\hat{g}_{m,k}$ denotes the channel that links the $m$-th AP to the $k$-th user. Then, we have
\begin{equation}
\hat{g}_{m,k}=\sqrt{\zeta_{m,k}}\left(\sqrt{1+\sigma_e^2}h_{m,k}-\sigma_e\tilde{h}_{m,k}\right),
\end{equation}
where $\zeta_{m,k}$ represents the slow fading coefficient,$h_{m,k}$ stands for the small scale fading coefficient, $\tilde{h}_{m,k}$ is the error in the channel estimate which follows a complex normal distribution with zero mean and unit variance and $\sigma_e^2$ can be interpreted as the quality of the channel estimate. It follows that 
\begin{equation}
    g_{m,k}=\frac{1}{\tau}\left(\hat{g}_{m,k}+\tilde{g}_{m,k}\right),
\end{equation}
with $\tau=\sqrt{1+\sigma_e^2}$, and $\tilde{g}_{m,k}=\sigma_e^2\tilde{h}_{m,k}$.

The  channel and the channel estimate are related  by  
\begin{equation}
    \mathbf{G}^H=\frac{1}{\tau}\left(\hat{\mathbf{G}}^H+\tilde{\mathbf{G}}^H\right),
\end{equation}
\begin{equation}
    \mathbf{G}^{\left(o\right)^H}=\frac{1}{\tau}\left(\hat{\mathbf{G}}^{\left(o\right)^H}+\tilde{\mathbf{G}}^{\left(o\right)^H}\right),
\end{equation}

\begin{subsection}{AP clustering}
Users are separated into disjoint clusters based on the largest large-scale fading coefficient, where the set $\mathcal{K}_i$ contains the index of the users that belong to cluster $i$. Similarly, the APs that provide service to the users in cluster $i$ are gathered in set $\mathcal{M}_i$. We can then define the effective channel matrix as follows:
\begin{align}
    g_{m,k}=\begin{cases}
        \hat{g}_{m,k},& m\in\mathcal{M}_i \text{ and } k\in\mathcal{K}_i\\
        0, &\text{Otherwise}.
    \end{cases}
\end{align}
At the receiver, we obtain
\begin{equation}
    \mathbf{y}=\frac{1}{\tau}\left(\hat{\mathbf{G}}^H+\tilde{\mathbf{G}}^H\right)\mathbf{x}+\frac{1}{\tau}\left(\hat{\mathbf{G}}^{\left(o\right)^H}+\tilde{\mathbf{G}}^{\left(o\right)^H}\right)\mathbf{x}+\mathbf{n},
\end{equation}
Thus, we can identify the terms related to the multiuser interference (MUI) and the noise in the following equation:
\begin{align}
  \mathbf{y}=&\underbrace{\frac{1}{\tau}\hat{\mathbf{G}}^H\mathbf{P}\mathbf{s}}_{\text{MUI Cancelled by } \mathbf{P}}+\underbrace{\frac{1}{\tau}\left(\hat{\mathbf{G}}^{\left(o\right)^H}+\tilde{\mathbf{G}}^{\left(o\right)^H}\right)\mathbf{P}\mathbf{s}}_{\text{MUI $\rightarrow$ OCL}}\nonumber\\
  &+\underbrace{\frac{1}{\tau}\tilde{\mathbf{G}}^H\mathbf{P}\mathbf{s}}_{\text{MUI $\rightarrow$ imperfect CSI}}+\underbrace{\mathbf{n}}_{\text{noise}}.\label{interference in the received signal}
\end{align}
\end{subsection}

\subsection{ICL and OCL Interference}

Equation \eqref{interference in the received signal} shows that the received signal is not only corrupted by noise, but also by interference emerging from two different sources. The ICL interference is produced by users inside the same cluster. In general, a precoder is implemented to cope with MUI inside the cluster. Nonetheless, imperfect channel state information (CSI) prevents the precoder from operating at its full potential, resulting in residual MUI. In other words, residual MUI is a consequence of the error in the channel estimate $\tilde{\mathbf{G}}^H$. For the uplink, receive processing approaches \cite{jidf,spa,mfsic, mbdf} have to be modified for this purpose.

On the other hand, users and APs outside the cluster produce OCL interference. Clusters are formed by employing the largest large-scale fading coefficient. The rationale is that two different APs that are far away from each other should belong to different clusters, so that OCL interference is mitigated. However, the interference arriving from neighboring clusters is detrimental to the overall performance. This interference is related to the channel matrix $\mathbf{G}^{\left(o\right)^H}$.

Both the ICL and OCL interference can degrade the performance heavily. This is particularly true in the high SNR regime, since increasing the transmitted power yields an increase in the power of the interference. Therefore, robust techniques, capable of dealing with these two sources of interference are crucial to achieving the potential of CF-MIMO systems.

\section{Proposed Robust MMSE precoder}

Robust precoding and beamforming approaches have been reported in the last two decades, with applications to multiple-antenna systems \cite{rmmseprec,rmmsecf,locsme,okspme,lrcc,baplnc,mbthp,rmbthp}. The proposed RMMSE precoder must minimize the effect of imperfect CSI, ICL and OCL interference. By letting $\mathbb{E}\left[\lVert\frac{1}{\tau}{\mathbf{G}}^{\left(o\right)^H}\mathbf{P}\mathbf{s}\rVert^2\right]\to 0$ the precoder minimizes the effects of ICL and OCL interference and performs as close as possible to the case where OCL interference is perfectly suppressed. To obtain such a precoder, we incorporate a penalty function into the objective function, and then solve the optimization problem:
    \begin{align}
    \left\{\mathbf{P},f\right\}=& \text{argmin}~~\underbrace{\mathbb{E}\left[\lVert\mathbf{s}-f^{-1}\mathbf{y}'\rVert^2\right]}_{T_1}+\underbrace{\mathbb{E}\left[\lVert\frac{1}{\tau}\mathbf{G}^{\left(o\right)^H}\mathbf{P}\mathbf{s}\rVert^2\right]}_{T_2}\nonumber\\    
    &\text{subject to}~~\mathbb{E}\left[\lVert\mathbf{x}\rVert^2\right]=\text{tr}\left(\mathbf{P}\mathbf{P}^H\right)=P_t,
\end{align}where $\mathbf{y}'=\frac{1}{\tau}\hat{\mathbf{G}}^H\mathbf{P}\mathbf{s}+\mathbf{n}$, and $f$ is a normalization factor.

We begin the derivation by expanding $T_{1}$ as follows:
\begin{equation}
    T_{
    1}=\mathbb{E}\left[\mathbf{s}^H\mathbf{s}\right]-f^{-1}\mathbb{E}\left[\mathbf{s}^H\mathbf{y'}\right]-f^{-1}\mathbb{E}\left[\mathbf{y'}^{H}\mathbf{s}\right]+f^{-2}\mathbb{E}\left[\mathbf{y'}^H\mathbf{y'}\right]\label{T1 expected values}
\end{equation}
By evaluating the expected values in \eqref{T1 expected values}, we obtain   $\mathbb{E}\left[\mathbf{s}^H\mathbf{s}\right]=K$, $  \mathbb{E}\left[\mathbf{s}^H\mathbf{y}'\right]=\frac{1}{\tau}\text{tr}\left(\hat{\mathbf{G}}^{H}\mathbf{P}\right)$, $\mathbb{E}\left[\mathbf{y'}^H\mathbf{s}\right]=\frac{1}{\tau}\text{tr}\left(\mathbf{P}^{H}\hat{\mathbf{G}}\right)$, and
$\mathbb{E}\left[\mathbf{y'}^H\mathbf{y'}\right]=\frac{1}{\tau^2}\text{tr}\left(\mathbf{P}\mathbf{P}^H\hat{\mathbf{G}}\hat{\mathbf{G}}^H\right)+K\sigma_n^2$. Additionally, we have
\begin{equation}
T_2  =\frac{1}{\tau^2}\text{tr}\left(\boldsymbol{\Psi}\mathbf{P}\mathbf{P}^H\right),
\end{equation}
where $\Psi=\mathbb{E}\left[\mathbf{G}^{\left(o\right)}\mathbf{G}^{\left(o\right)^H}\right]$.
It follows that the optimization problem can be reformulated as 
\begin{align}
    \left\{\mathbf{P},f\right\}=& \text{argmin}~~J_{I}\nonumber\\  
     \hspace{-1em} \text{subject to}~~\mathbb{E}&\left[\lVert\mathbf{x}\rVert^2\right]=P_t,\label{minimization problem modified}
\end{align}
where
\begin{align}
    J_{I}=& K-\frac{f^{-1}}{\tau}\text{tr}\left(\mathbf{P}^H\hat{\mathbf{G}}\right)-\frac{f^{-1}}{\tau}\text{tr}\left(\hat{\mathbf{G}}^H\mathbf{P}\right)+f^{-2}K\sigma_n^2\nonumber\\
    &+\frac{f^{-2}}{\tau^2}\text{tr}\left(\mathbf{P}\mathbf{P}^H\hat{\mathbf{G}}\hat{\mathbf{G}}^H\right)+\frac{1}{\tau^2}\text{tr}\left(\boldsymbol{\Psi}\mathbf{P}\mathbf{P}^H\right).    
\end{align}
The Lagrangian function of the optimization problem is given by 
\begin{align}
    \mathcal{L}(\mathbf{P},f,\lambda)=&J_I+\lambda\left[\text{tr}\left(\mathbf{P}\mathbf{P}^H\right)-P_t\right].\label{Lagrangian proposed precoder}
\end{align}
The partial derivatives of the Lagrangian function are given by
\begin{align}
\frac{\partial\mathcal{L}\left(\mathbf{P},f,\lambda\right)}{\partial\mathbf{P}^{*}}=&-\frac{f^{-1}}{\tau}\hat{\mathbf{G}}+\frac{f^{-2}}{\tau^2}\hat{\mathbf{G}}\hat{\mathbf{G}}^{H}\mathbf{P}+\frac{1}{\tau^{2}}\boldsymbol{\Psi}\mathbf{P}+\lambda\mathbf{P},\label{partial derivative 1}
\end{align}
\begin{align}
\frac{\partial\mathcal{L}\left(\mathbf{P},f,\lambda\right)}{\partial f}=&\frac{f^{-2}}{\tau}\left(\text{tr}\left(\hat{\mathbf{G}}^{H}\mathbf{P}\right)+\text{tr}\left(\mathbf{P}^{H}\hat{\mathbf{G}}\right)\right)\nonumber\\
    &-\frac{2f^{-3}}{\tau^2}\text{tr}\left(\mathbf{P}\mathbf{P}^{H}\hat{\mathbf{G}}\hat{\mathbf{G}}^{H}\right)-2f^{-3}K\sigma_n^2.\label{partial derivative 2}
\end{align}
Equating \eqref{partial derivative 1} and \eqref{partial derivative 2} to zero and rearranging terms, we obtain

\begin{equation}
    f\tau\hat{\mathbf{G}}=\left[\hat{\mathbf{G}}\hat{\mathbf{G}}^H+f^2\boldsymbol{\Psi}+f^2\tau^2\lambda\mathbf{I}\right]\mathbf{P},\label{partial derivative 1 solved}
\end{equation}

\begin{equation}
    f\tau\text{tr}\left(\hat{\mathbf{G}}\mathbf{P}^H\right)=\text{tr}\left(\mathbf{P}\mathbf{P}^{H}\hat{\mathbf{G}}\hat{\mathbf{G}}^{H}\right)+\tau^2K\sigma_n^2.\label{partial derivative 2 solved}
\end{equation}

Multiplying \eqref{partial derivative 1 solved} by $\mathbf{P}^H$ to the right-hand side and taking the trace, we obtain
\begin{align}
    f\tau\text{tr}\left(\hat{\mathbf{G}}\mathbf{P}^H\right)=&\text{tr}\left(\mathbf{P}\mathbf{P}^{H}\hat{\mathbf{G}}\hat{\mathbf{G}}^{H}\right)+f^2\text{tr}\left(\mathbf{P}\mathbf{P}^H\boldsymbol{\Psi}\right)\nonumber\\  &+f^2\tau^2\lambda\text{tr}\left(\mathbf{P}\mathbf{P}^H\right).\label{partial derivative 1 trace}
\end{align}

Employing \eqref{partial derivative 1 trace} and \eqref{partial derivative 2 solved}, we get

\begin{equation}
    \tau^2K\sigma_n^2=f^2\text{tr}\left(\mathbf{P}\mathbf{P}^H\boldsymbol{\Psi}\right)+f^2\tau^2\lambda\text{tr}\left(\mathbf{P}\mathbf{P}^H\right).
\end{equation}
Then, we have
\begin{equation}
    \lambda\text{tr}\left(\mathbf{P}\mathbf{P}^H\right)=\frac{K\sigma_n^2}{f^2}-\frac{\text{tr}\left(\mathbf{P}\mathbf{P}^H\boldsymbol{\Psi}\right)}{\tau^2}.\label{lambda not isolated}
\end{equation}
Using the total power constraint $\text{tr}\left(\mathbf{P}\mathbf{P}^H\right)=P_t$ in \eqref{lambda not isolated} we obtain the expression for $\lambda$:
\begin{equation}
    \lambda=\frac{K\sigma_n^2}{f^2P_t}-\frac{\text{tr}\left(\mathbf{P}\mathbf{P}^H\boldsymbol{\Psi}\right)}{\tau^2P_t}.
\end{equation}

From the partial derivatives, we also obtain
\begin{equation}    
\mathbf{P}=f\tau{\bar{\mathbf{P}}},
\end{equation}
where
\vspace{-0.5em}
\begin{equation}
{\bar{\mathbf{P}}}=(\underbrace{\hat{\mathbf{G}}\hat{\mathbf{G}}^H+f^2\boldsymbol{\Psi}+\lambda f^2\tau^2\mathbf{I}}_{\mathbf{P}^{\left(i\right)}})^{-1}\hat{\mathbf{G}},\label{p_bar robust mmse}
\end{equation}
    \begin{equation}
    f=\frac{1}{\tau}\sqrt{\frac{P_t}{\text{tr}\left(\bar{\mathbf{P}}\bar{\mathbf{P}}^{H}\right)}},\label{power scaling factor}
\end{equation}
assuming that the inverse of $\mathbf{P}^{\left(i\right)}$ exists.

Note that $\mathbf{P}$ depends on $\lambda$ and vice-versa. Therefore, we employ an alternating optimization (AO) framework, where one of the variables is fixed while the value of the other variable, which minimizes $J_{I}$ is computed. The standard MMSE precoder can be used as the initial state. Then, we update the parameter $\lambda$ iteratively. 

\section{Sum-Rate Analysis}
Let us denote the received signal at user $k$, which belongs to cluster $i$ by $y_{k,i}$ By taking the expected value of the $l_2$-norm of $y_{k,i}$ and using \eqref{receive vector at cluster i complete}, we obtain
\begin{align}
    \mathbb{E}\left[\lvert y_{k,i}\rvert^2\right]=\frac{1}{\tau^2}\lvert\hat{\mathbf{g}}_{k}^{H}\mathbf{p}_k\rvert^2+\frac{1}{\tau^2}\delta_k+\phi_k+\sum_{j=1,j\neq k}^K\lvert\hat{\mathbf{g}}_{k}^H\mathbf{p}_j\rvert^2+\sigma_n^2,
\end{align}
 where $\delta_k=\sum_{t=1}^{K}\Re\{\left(\hat{\mathbf{g}}^H_{k}\mathbf{p}_t\right)\left(\tilde{\mathbf{g}}^H_{k}\mathbf{p}_t\right)\}+\sum_{l=1}^K\lvert\tilde{\mathbf{g}}_{k}^H\mathbf{p}_l\rvert^2$ denotes the power of the interference generated by the imperfect CSI, and $\phi_k=\sum_{q=1}^K\lvert\mathbf{g}^{\left(o\right)^H}_{k}\mathbf{p}_q\rvert^2$ denotes the power of the OCL interference. Then, 
the signal to interference-plus-noise ratio at the $k$-th user is given by
\begin{equation}
    \gamma_k=\frac{\frac{1}{\tau^2}\lvert\hat{\mathbf{g}}_{k}^H\mathbf{p}_k\rvert^2}{\frac{1}{\tau^2}\delta_k+\phi_k+\sum_{j=1,j\neq k}^K\lvert\hat{\mathbf{g}}_{k}^H\mathbf{p}_j\rvert^2+\sigma_n^2},
\end{equation}
Thus, the instantaneous rate at the $k$-th user considering Gaussian signaling is given by $R_k=\log_2(1+\gamma_k)$. Due to the imperfect CSI, the instantaneous rates are not achievable. Therefore, we employ the ergodic sum-rate (ESR), which can be computed by
\begin{equation}
    S_r=\sum_{k=1}^K\mathbb{E}\left[R_k\right].
\end{equation}


\section{Simulation Results}
We evaluate the performance of the proposed RMMSE precoders with OCL interference suppression (RMMSE-OCLIS) and with perfect OCL interference suppresion (RMMSE-pOCLIS) via numerical examples and compare them with the conventional MMSE network-wide (MMSE-NW) technique which has knowledge of all channel coefficients. Additionally, we also consider the conventional MMSE precoder, which has knowledge of the ICL interference and no knowledge of the OCL interference and, therefore, does not perform OCL interference suppression. We consider a CF-MIMO system where $24$ APs were deployed. The APs provide service to six users distributed randomly over the area of interest. The users and the APs are split into three disjoint clusters. To compute the ergodic sum-rate, a total of $10~000$ trials were considered. The large-scale fading coefficients are defined by
\begin{equation}
     \zeta_{m,k}=P_{m,k}\cdot 10^{\frac{\sigma^{\left(\textrm{s}\right)}z_{m,k}}{10}},
\end{equation}
where $P_{m,k}$ represents the path loss. The log-normal shadowing is modeled by  $10^{\frac{\sigma^{\left(\textrm{s}\right)}z_{m,k}}{10}}$, where $\sigma^{\left(\textrm{s}\right)}=8$ dB is the standard deviation and the random variable $z_{m,k}$ is Gaussian distributed with zero mean and unit variance. The path loss in dB is calculated using a three-slope model as \par\noindent\small
 \begin{align}
     P_{m,k}=\begin{cases}
  -L-35\log_{10}\left(d_{m,k}\right), & \text{$d_{m,k}>d_1$} \\
  -L-15\log_{10}\left(d_1\right)-20\log_{10}\left(d_{m,k}\right), & \text{$d_0< d_{m,k}\leq d_1$}\\
    -L-15\log_{10}\left(d_1\right)-20\log_{10}\left(d_0\right), & \text{otherwise,}
\end{cases}
 \end{align}\normalsize
 where $d_{m,k}$ is the distance between the $m$-th AP and $k$-th users, $d_1=50$ m, $d_0= 10$ m, and the attenuation $L$ is \par\noindent\small
 \begin{align}
L=&46.3+33.9\log_{10}\left(f_c\right)-13.82\log_{10}\left(h_{\textrm{AP}}\right)\nonumber\\
&-\left(1.1\log_{10}\left(f_c\right)-0.7\right)h_u+\left(1.56\log_{10}\left(f_c\right)-0.8\right),
 \end{align}\normalsize
 where $h_{\textrm{AP}}=11.65$ m and $h_{u}=1.65$ are the positions of, respectively, the AP and the user equipment above the ground and frequency $f_c= 1900$ MHz. The noise variance is $\sigma_w^2=T_o k_B B N_f,$ where $T_o=290$ K is the noise temperature, $k_B=1.381\times 10^{-23}$ J/K is the Boltzmann constant, $B=20$ MHz is the bandwidth and $N_f=9$ dB is the noise figure. 

\begin{figure}[h]
    \begin{center} \includegraphics[width=0.85\linewidth]{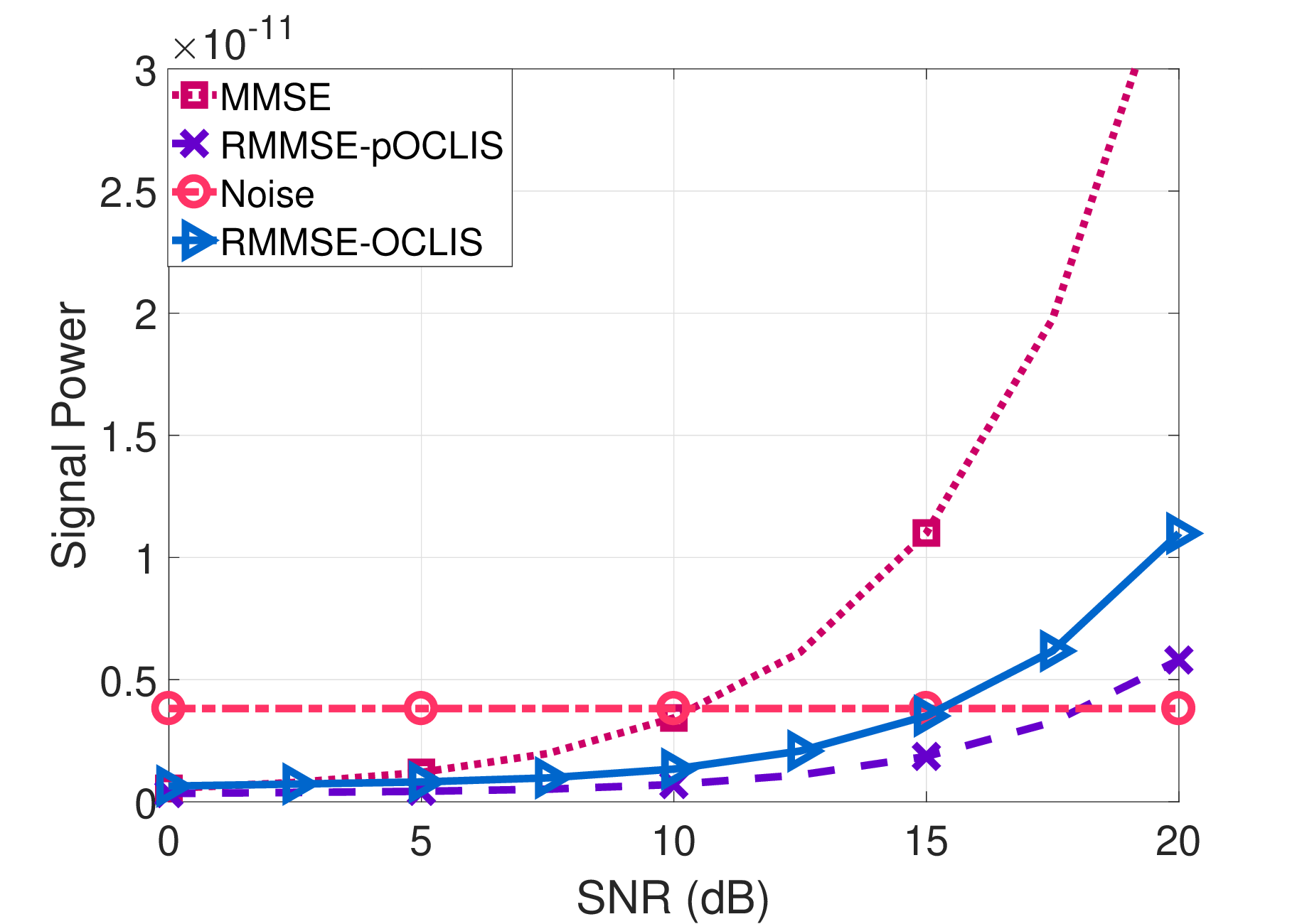}
    \vspace{-1.25em}    
    \caption{Power of the OCL interference considering different mitigation schemes.}
    \label{fig:Fig1}
    \end{center}
\end{figure}

In the first example, we analyze the power of the OCL interference. Specifically, Fig. \ref{fig:Fig1}  shows the OCL interference power considering a conventional MMSE precoder (without OCL interference mitigation),  the proposed RMMSE-OCLIS and  RMMSE-pOCLIS precoders, i.e., completely removing the interference caused by $\mathbf{G}^{(o)^H}$. Moreover, the power of the noise is included for comparison. Note that perfect OCL interference mitigation reduces the power of the interference almost to the noise level at 20 dB. However, this requires the knowledge of the channel coefficients of the APs outside the cluster, incurring high signaling load and computational complexity. In contrast, the proposed RMMSE precoder with statistical OCL interference shows an efficient performance by not requiring exact knowledge of the channel coefficients while greatly reducing the power of the interference. 

\begin{figure}[h]
    \begin{center} \includegraphics[width=0.85\linewidth]{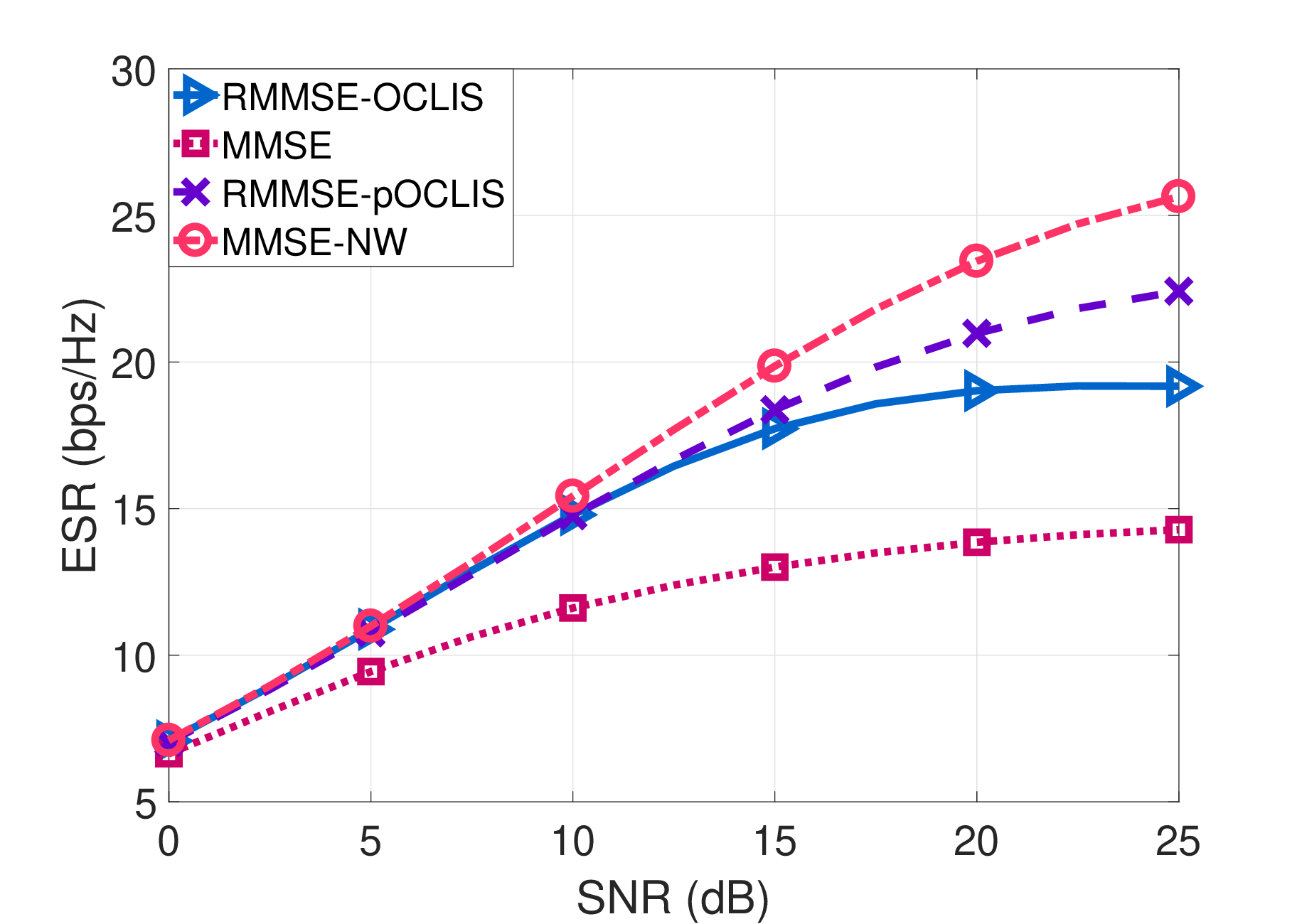}
        \vspace{-1.25em}
    \caption{ESR performance of the precoders with different OCL interference mitigation schemes}
    \label{Fig2}
    \end{center}
\end{figure}

In the second example, we assess the ESR performance of the proposed techniques in Fig. \ref{Fig2}. The MMSE-NW and the RMMSE-pOCLIS precoders obtain the best result. However, they are not practical since MMSE-NW is not scalable and the proposed RMMSE-pOCLIS precoder requires perfect CSI of the APs outside the cluster. In contrast, the proposed RMMSE-OCLIS is promising since it does not require perfect CSI of the APs outside the cluster and greatly enhances the sum-rate performance when compared to the MMSE precoder without OCL interference.

\section{Conclusions}
In this paper, a robust MMSE precoder for imperfect CSI, ICL and OCL mitigation has been developed for CF-MIMO systems. The proposed precoder improves the overall performance of the system by effectively mitigating the OCL interference. In contrast to network-wide approaches, the proposed technique has a low-signaling load, being suitable for practical systems.

\bibliographystyle{IEEEbib}
\bibliography{CellFree}

\end{document}